\documentclass[reprint, amsmath, amssymb, aps, prb, superscriptaddress]{revtex4-2}
\usepackage{amsmath}
\usepackage{amssymb}
\usepackage{graphicx}
\usepackage{epstopdf}
\usepackage{natbib}
\usepackage{array}
\usepackage[utf8]{inputenc}
\usepackage{xspace}
\usepackage{multirow}
\usepackage{changes}
\usepackage{dcolumn}
\usepackage{bm}
\usepackage{braket}

\newcommand{\smo}{SrMoO$_3$}

\newcommand{\svo}{SrVO$_3$}
\newcommand{\Aunit}{$\mu\Omega\,\text{cm}/\text{K}^2$}

\RequirePackage[
  hyperindex,colorlinks,bookmarksnumbered,
  plainpages=true,pdfstartview=FitH]{hyperref}
\hypersetup{urlcolor={blue!50!black}, linkcolor={red!50!black},citecolor={blue!50!black}} 
\usepackage{hyperref}
\usepackage[all]{hypcap}

\makeatletter
\def\maketitle{
\@author@finish
\title@column\titleblock@produce
\suppressfloats[t]}
\makeatother

\begin{document}
\title{Fermi-liquid $T^2$ resistivity: Dynamical mean-field theory meets experiment}
\author{Fabian B.~Kugler}
\thanks{These authors contributed equally to this work.}
\affiliation{Center for Computational Quantum Physics,
             Flatiron Institute,
             162 5th Avenue, New York, New York 10010, USA}
\affiliation{Institute for Theoretical Physics, University of Cologne, 50937 Cologne, Germany}
\author{Jeremy Lee-Hand}
\thanks{These authors contributed equally to this work.}
\affiliation{Department of Physics and Astronomy,
             Stony Brook University,
             Stony Brook, New York, 11794-3800, USA}
\author{Harrison LaBollita}
\thanks{These authors contributed equally to this work.}
\affiliation{Center for Computational Quantum Physics,
             Flatiron Institute,
             162 5th Avenue, New York, New York 10010, USA}
\author{Lorenzo Van Mu\~{n}oz}
\affiliation{Department of Physics,
             Massachusetts Institute of Technology,
             77 Massachusetts Avenue, Cambridge, Massachusetts 02139, USA}
\author{Jason Kaye}
\affiliation{Center for Computational Quantum Physics, Flatiron Institute, 162 5th Avenue, New York, New York 10010, USA}
\affiliation{Center for Computational Mathematics, Flatiron Institute, 162 5th Avenue, New York, New York 10010, USA}
\author{Sophie Beck}
\affiliation{Center for Computational Quantum Physics,
             Flatiron Institute,
             162 5th Avenue, New York, New York 10010, USA}
\author{Alexander Hampel}
\affiliation{Center for Computational Quantum Physics,
             Flatiron Institute,
             162 5th Avenue, New York, New York 10010, USA}
\author{Antoine Georges}
\affiliation{Coll{\`e}ge de France, 11 Place Marcelin Berthelot, 75005 Paris, France}
\affiliation{Center for Computational Quantum Physics,
             Flatiron Institute,
             162 5th Avenue, New York, New York 10010, USA}
\affiliation{CPHT, CNRS, {\'E}cole Polytechnique, IP Paris, F-91128 Palaiseau, France}
\affiliation{DQMP, Universit{\'e} de Gen{\`e}ve, 24 Quai Ernest Ansermet, CH-1211 Gen{\`e}ve, Suisse}
\author{Cyrus E. Dreyer}
\thanks{Contact author, cyrus.dreyer@stonybrook.edu}
\affiliation{Department of Physics and Astronomy,
             Stony Brook University,
             Stony Brook, New York, 11794-3800, USA}
\affiliation{Center for Computational Quantum Physics,
             Flatiron Institute,
             162 5th Avenue, New York, New York 10010, USA}
\date{\today}

\begin{abstract}
Direct-current resistivity is a key probe for the physical properties of materials. In metals, Fermi liquid (FL) theory serves as the basis for understanding transport. A $T^2$ behavior of the resistivity is often taken as a signature of FL electron-electron scattering. However, the presence of impurity and phonon scattering as well as material-specific aspects such as Fermi surface geometry can complicate this interpretation. We demonstrate how density-functional theory combined with dynamical mean-field theory can be used to elucidate the FL regime. We take as examples SrVO$_{3}$ and SrMoO$_{3}$, two moderately correlated perovskite oxides, and establish a precise framework to analyze the FL behavior of the self-energy at low energy and temperature. Reviewing published low-temperature resistivity measurements, we find agreement between our calculations and experiments performed on samples with exceptionally low residual resistivity. This comparison emphasizes the need for further theoretical, synthesis, and characterization developments in these and other FL materials. 
\end{abstract}

\maketitle

The direct-current (dc) resistivity as a function of temperature, $\rho(T)$, is a fundamental probe in solid-state physics. In metals, $\rho(T)$ typically increases as a power law of $T$, the specific form of which in principle differentiates between distinct scattering mechanisms. A cornerstone of such an analysis is Fermi liquid (FL) theory~\cite{Landau:1956}, which provides a robust framework for understanding interacting fermions. A key prediction of the theory is the $\rho\propto T^{2}$ scaling originating from electron-electron (el-el) Umklapp scattering \cite{Landau1937} (for a recent discussion, see \cite{Behnia2022}).

In most materials that do not have especially strong electronic correlations, electron-phonon (el-ph) scattering dominates $\rho(T)$ except possibly at very low $T$ \cite{Ziman1960,Bass1990}. Compared to this el-ph case \cite{Giustino2017}, there has been considerably 
less work \cite{Abramovitch2023,Swift2017,Deng2016,Pourovskii2017} 
analyzing the regime of transport dominated by el-el scattering.
Indeed, the low temperatures necessary to resolve this regime poses severe challenges both to theory, where accurate low-energy scattering rates are difficult to obtain, and to experiment, where the FL resistivity must be distinguished from the residual resistivity induced by impurities and disorder.

It is crucial to develop and validate FL el-el scattering rates in moderately correlated materials for several reasons. First, it provides a stringent test of the computational methods; since the magnitude of scattering rates are so low, rigorous convergence and careful extraction of parameters is required. Next, there are numerous reports of $T^2$ scaling of resistivity in regimes where it is not expected \cite{Lin2015_2,Behnia2022,Inoue1998}. For example, some materials with moderate correlations (indicated by low resistivities) exhibit $\rho \propto T^2$ up to temperatures approaching room temperature (RT); FL el-el scattering is often invoked in these cases, and accurate values are necessary to confirm or disprove this scattering mechanism. Finally, to determine the applicability of common approximations (e.g., assuming a local self energy, neglecting vertex corrections) in real materials, it is necessary to differentiate shortcomings of the actual transport theory from uncertainties associated with the computational procedure or experimental benchmark data.

In this work, we overcome the computational limitations to low-temperature dc resistivity, allowing us to perform a quantitative theory-experiment comparison for FL el-el scattering using two systems: the perovskite oxides \svo{} and \smo{}. These materials exhibit moderate correlations
and relatively low values of $\rho$ (in fact, \smo{} has the lowest reported RT resistivity of any perovskite oxide \cite{Nagai2005}). 
For both materials (\svo{} in particular), there has been significant experimental study of the dc resistivity \cite{Nozaki1991,Itoh1991,lan2003,Ahn2022,xu2019,Roth2021,Fouchet2016,Brahlek2015,Brahlek2024,Shoham2020,Zhang2016,Mirjolet2019,mirjolet2021,Mirjolet2021_2,Reyes2000,Inoue1998,berry2022,Wang2001,Cappelli2022,Lekshmi2005,Radetinac2016,Nagai2005}.
A $T^2$ dependence was observed up to hundreds of Kelvin
and often attributed to FL el-el scattering, though this interpretation
has been debated \cite{Inoue1998,Dylla2019,mirjolet2021,Abramovitch2024}.

We first perform a critical review of the experimental landscape for the materials. Then, we show that leveraging recent algorithmic advances and careful analysis of the numerical data allows density functional theory combined with dynamical mean-field theory (DFT+DMFT) \cite{Georges:1996,Kotliar:2006} to successfully elucidate the low-$T$ FL behavior. In particular, our calculations are consistent with a low-$T$ ($\lesssim 30$ K) $\rho \propto T^2$ regime that is only resolvable in experiments with ultralow residual resistivity and is distinct from the $T^2$ dependence observed at higher $T$. Finally, we discuss fundamental questions on the theory of transport raised by our findings.

\begin{figure}
\includegraphics[width=\linewidth]{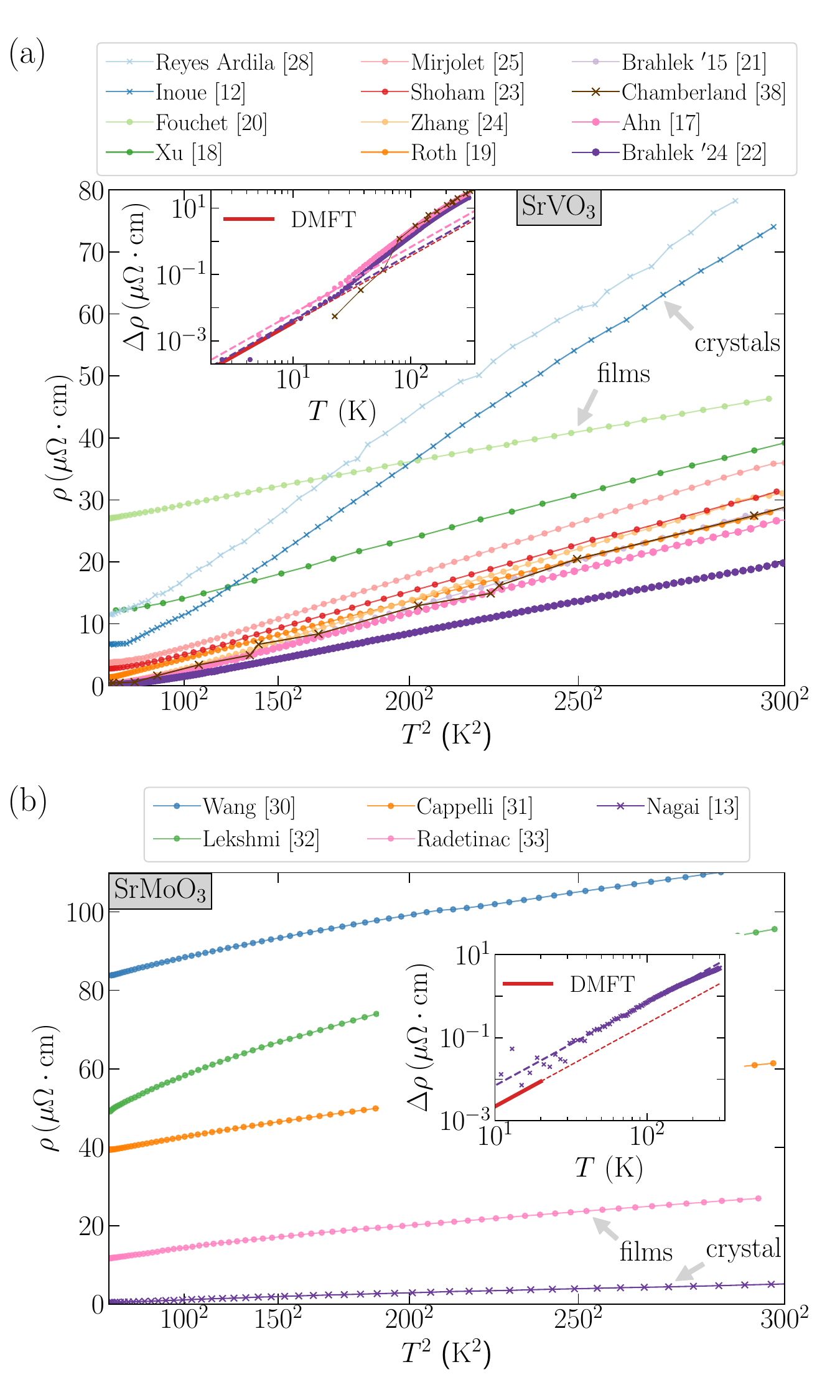}
\caption{Resistivity measurements versus $T^2$ on a linear scale for (a) \svo{} and (b) \smo{}, extracted from Refs.~\citenum{Chamberland1971,Reyes2000,Inoue1998,Fouchet2016,xu2019,mirjolet2021,Shoham2020,Roth2021,Zhang2016,Brahlek2015,Ahn2022,Brahlek2024,Wang2001,Cappelli2022,Lekshmi2005,Radetinac2016,Nagai2005}. The inset in (a) shows $\Delta \rho = \rho - \rho_0$, with $\rho_0$ the $T=0$ residual resistivity, for the data from Ahn \textit{et al.}~\cite{Ahn2022} and Brahlek \textit{et al.}~\cite{Brahlek2024}, clearly showing two different $T^2$ regimes above and below $\sim 20$~K. The inset in (b) shows the same for the data from Nagai \textit{et al.}~\cite{Nagai2005}.}
\label{fig:exp}
\end{figure}

The experimental situation in terms of dc resistivities of \svo{} and \smo{} is summarized in Fig.~\ref{fig:exp},  where we plot $\rho$ versus $T^2$ extracted from several selected \footnote{Where data for multiple samples are presented, we take the highest RRR sample. For \svo{}, we exclude previous powder results \cite{Nozaki1991,Itoh1991,lan2003} and the recent single-crystal results of Ref.~\citenum{berry2022} due to the very large $\rho_0$ (though Ref.~\citenum{berry2022} is included in Table~S1. We also exclude Ref.~\citenum{Giannakopoulou1995} due to difficulties in extracting the data from the plots and Ref.~\citenum{Maekawa2005} due to lacking low-$T$ measurements. The data for Mirjolet \textit{et al.} are from Ref.~\citenum{mirjolet2021}.}
experimental works differentiated by single-crystal samples versus epitaxial films. For both materials, the curves appear to follow a $\rho(T)=\rho_0+AT^2$ trend up to RT. 
Interestingly, for \svo{}, most of the reported single-crystal data \cite{Reyes2000,Inoue1998,berry2022} have higher $A$ coefficients and higher residual resistivities $\rho_0$ than the majority of the thin films \cite{Ahn2022,xu2019,Roth2021,Fouchet2016,Brahlek2015,Brahlek2024,Shoham2020,Zhang2016,Mirjolet2019,mirjolet2021,Mirjolet2021_2}; the one exception is the single-crystal results of Ref.~\citenum{Chamberland1971}, which actually agree with the trends of the thin films. For \smo{}, the lone single-crystal result \cite{Nagai2005} has by far the lowest $A$ and $\rho_0$ compared to films \cite{Wang2001,Lekshmi2005,Radetinac2016,Cappelli2022}.

In the low-$\rho_0$ films of \svo{} \cite{Ahn2022,Brahlek2024}, there are actually \emph{two} regimes of approximately $T^2$ behavior, separated by a crossover around 20--30 K \cite{Brahlek2024}. This is clearly seen in the inset of Fig.~\ref{fig:exp}(a), showing $\Delta\rho=\rho-\rho_0$ from Refs.~\citenum{Ahn2022} and \citenum{Brahlek2024} on a log-log scale. This effect was attributed \cite{Abramovitch2024} to a crossover between el-el- and el-ph-dominated scattering and is difficult to resolve in samples with higher $\rho_0$. The single-crystal data from Ref.~\citenum{Chamberland1971} also appears to have a second, low-$T$ $T^2$ regime, though the data at low $T$ is not sufficient to determine whether the coefficient quantitatively agrees with Refs.~\citenum{Ahn2022} and \citenum{Brahlek2024}. For \smo{}, though the low-$T$ resolution of Ref.~\citenum{Nagai2005} is more limited, it appears that the data similarly lays below the high-$T$ $T^2$ trend, see inset of Fig.~\ref{fig:exp}(b).

A quantitative analysis of the experimental $\rho(T)$ is given in Table~S1 of the Supplemental Material \cite{SM} (SM), see also Refs.~\citenum{berry2022,Reyes2000,Inoue1998,Chamberland1971,Fouchet2016,xu2019,mirjolet2021,Shoham2020,Roth2021,Zhang2016,Brahlek2015,Ahn2022,Brahlek2024,Wang2001,Cappelli2022,Lekshmi2005,Radetinac2016,Nagai2005} therein. We fit the curves of Fig.~\ref{fig:exp} to extract $A$ in two regimes: (i) $T$ up to RT and (ii) $T$ up to $20$~K ($50$~K) for \svo{} (\smo{}). We also extract $\rho_0$ by extrapolating the curves to $T=0$~\footnote{This was done by fitting the digitized data to $\rho(T)=\rho_0+AT^2$ either below 20 K, or 50 K if the fits to 20 K could not be achieved.} and order them by their residual-resistivity ratios (RRRs). 

For \svo{}, consistent with Fig.~\ref{fig:exp}(a), one finds a clear differentiation between the single-crystal and film results; we focus on the films since they exhibit the lowest $\rho_0$'s and largest RRRs. The $A$ coefficients fit up to RT are in the range $2$--$4 \times 10^{-4}$ \Aunit{}, while the spread of $A$'s fit below $20$~K illustrates the challenges of resolving small $\rho$ in the presence of impurity scattering. For films with a larger RRR (and also lower $\rho_0$), $A$ is about an order of magnitude smaller than the values fit up to RT. As shown below, if we interpret this low-$T$ regime to be the one dominated by el-el scattering \cite{Brahlek2024,Abramovitch2024}, its $A$ coefficient agrees with our DFT+DMFT result within error bars. For \smo{}, the inset of Fig.~\ref{fig:exp}(b) indicates that the low-$T$ $T^2$ regime of $\rho$ in the high-quality (single-crystal) sample \cite{Nagai2005} is not fully resolved. Interestingly, the fact that data points below 30~K seem to fall below the RT $T^2$ behavior (potentially indicating a lower value of the $A$ coefficient at low $T$) is also consistent with our DFT+DMFT calculations.

We now turn to our computational framework. The low-energy models for \svo{} and \smo{} are derived from DFT by downfolding the Kohn--Sham bands onto Wannier-like $t_{2g}$ orbitals. The models are solved within single-site DMFT using both continuous-time quantum Monte Carlo (QMC) in the hybridization expansion (CT-HYB) as implemented in the TRIQS software~\cite{parcollet_triqs_2015, aichhorn_dfttools_2016, Seth2016274, Merkel2022} and the numerical renormalization group (NRG) \cite{Bulla2008} in an implementation~\cite{Kugler2019,Kugler2020,Kugler2022a,Kugler2024} based on the QSpace tensor library \cite{Weichselbaum2012a,Weichselbaum2012b,Weichselbaum2020}. For the QMC imaginary-frequency solver, Pad\'e analytic continuation (AC) is used to obtain low-energy real-frequency data. For the NRG real-frequency solver, using the recently developed symmetric improved estimator for the self-energy is crucial for fine low-energy resolution \cite{Kugler2022}. Further computational details are given in the SM~\cite{SM} (and Refs.~\citenum{Kresse:1993bz,Kresse:1996kl,Kresse:1999dk,wannier90,aichhorn_dfttools_2016, parcollet_triqs_2015, Seth2016274,Merkel2022,Lee2021,Lee2017,Lee2016,Weichselbaum2012a,Weichselbaum2012b,Weichselbaum2020,Lechermann2006,LeeHand2021,Hampel2021,CompanionPaperLong,Sekiyama2004,Pavarini2004,Nekrasov2006,Maiti2006,Kaufmann2021} therein) and in Ref.~\cite{CompanionPaperLong}.

We compute the conductivity within DMFT using the Kubo formula neglecting vertex corrections (consistent with the momentum independence of the DMFT self-energy and vertex \cite{Khurana1990,Uhrig1995}). For \svo{} and \smo{} in their cubic crystal settings, the $t_{2g}$ orbitals are degenerate and the self-energy is band independent, $\Sigma_{mm'}(\omega,T)=\delta_{mm'}\Sigma(\omega,T)$, which enables the following simplifications. We consider $2|\mathrm{Im}\Sigma(0,T)| \!\ll\! T \!\ll\! \epsilon_{\text{F}}$, where $\epsilon_{\text{F}}$ is the Fermi energy and $k_B \!=\! \hbar \!=\! e \!=\! 1$. As interband contributions are negligible in the dc limit, one then finds~\cite{Berthod2013,Georges2021}
\begin{equation}
\label{eq:rho_int}
\rho(T)=\left[\Phi(\epsilon_{\text{F}})\int^{+\infty}_{-\infty}d\omega \left(-\frac{df}{d\omega}\right)\frac{1}{2\vert\text{Im}\Sigma(\omega,T)\vert}\right]^{-1}
.
\end{equation}
Here, $f$ is the Fermi--Dirac distribution function and $\Phi_{\alpha\alpha'}(\epsilon_{\text{F}})=\delta_{\alpha\alpha'}\Phi(\epsilon_{\text{F}})$ the 
transport function at $\epsilon_{\text{F}}$,
\begin{equation}
\Phi_{\alpha\alpha'}(\epsilon_{\text{F}})=2 \sum_{m}\int_{\text{BZ}} \frac{d^dk}{(2\pi)^d} v^\alpha_m(\textbf{k})v^{\alpha'}_m(\textbf{k})\delta(\epsilon_{\text{F}}-\epsilon_{m\textbf{k}}),
\label{eq:Phi}
\end{equation}
where $d$ is the dimensionality, $v^\alpha_m(\textbf{k})$ the velocity of band $m$ and $k$-point \textbf{k} in direction $\alpha$, and $\epsilon_{m\textbf{k}}$ its energy. 
The transport function is obtained by iterative adaptive integration \footnote{Results are converged to four digits at $1\,\mathrm{meV}$ broadening for both the transport function and the density of states.} leveraging Wannier interpolation using the Julia library \textsc{AutoBZ.jl} \cite{Kaye2023,Van-Munoz/Beck/Kaye:2024}.
For a FL, the self-energy fulfills to leading order in $\omega$ and $T$ 
\begin{equation}
\label{eq:IMSig_FL}
\text{Im}\Sigma(\omega,T)
\simeq
-C (\omega^2+\pi^2 T^2).
\end{equation}
Plugging Eq.~(\ref{eq:IMSig_FL}) into Eq.~(\ref{eq:rho_int}) eventually gives the FL dc resistivity to leading order as 
\begin{equation}
\label{eq:rho_FL}
\rho(T)
\simeq
A T^2
, \qquad
A = 24C/\Phi(\epsilon_{\text{F}})
.
\end{equation}

In principle, it should be straightforward to extract $C$ from the DMFT self-energy $\text{Im}\Sigma$, for $T < T^{\text{FL}}$, where $T^{\text{FL}}$ denotes the scale below which FL behavior holds [i.e., Eq.~(\ref{eq:IMSig_FL}) applies in the absence of other sources of inelastic scattering than el-el]. However, the small scattering rates of \svo{} and \smo{} make this extraction a challenging task. Thus, a careful analysis of how to obtain $C$ is necessary to have confidence in the numerical precision of the $A$ coefficients. For a full discussion, see Ref.~\citenum{CompanionPaperLong}; here, we briefly summarize the methods we found most reliable and used in this work.

Since most QMC algorithms operate on the imaginary-frequency axis, it is common practice to extrapolate Im$\Sigma(i\omega\rightarrow0^+,T)$ and fit a quadratic function to several values of $T$. However, for \svo{} and \smo{}, the value of $C$ from this procedure is sensitive to the choice of numerical parameters \cite{CompanionPaperLong}. Performing AC would allow direct evaluation of Im$\Sigma(\omega=0,T)$, but QMC+Pad\'e results for $|\omega| \!<\! T$ depend sensitively on the AC parameters, particularly the value of $\eta$ when evaluating the Pad\'e approximant at $\omega + i \eta$ \cite{CompanionPaperLong}. NRG works with real frequencies, circumventing the need for AC, but the $|\omega| \!<\! T$ region is also less reliable than $|\omega| \!>\! T$ since the Wilson chain is effectively cut on the scale of $T$ \cite{Weichselbaum2007,Weichselbaum2012b,Lee2016}.

\begin{figure}
\includegraphics[width=\linewidth]{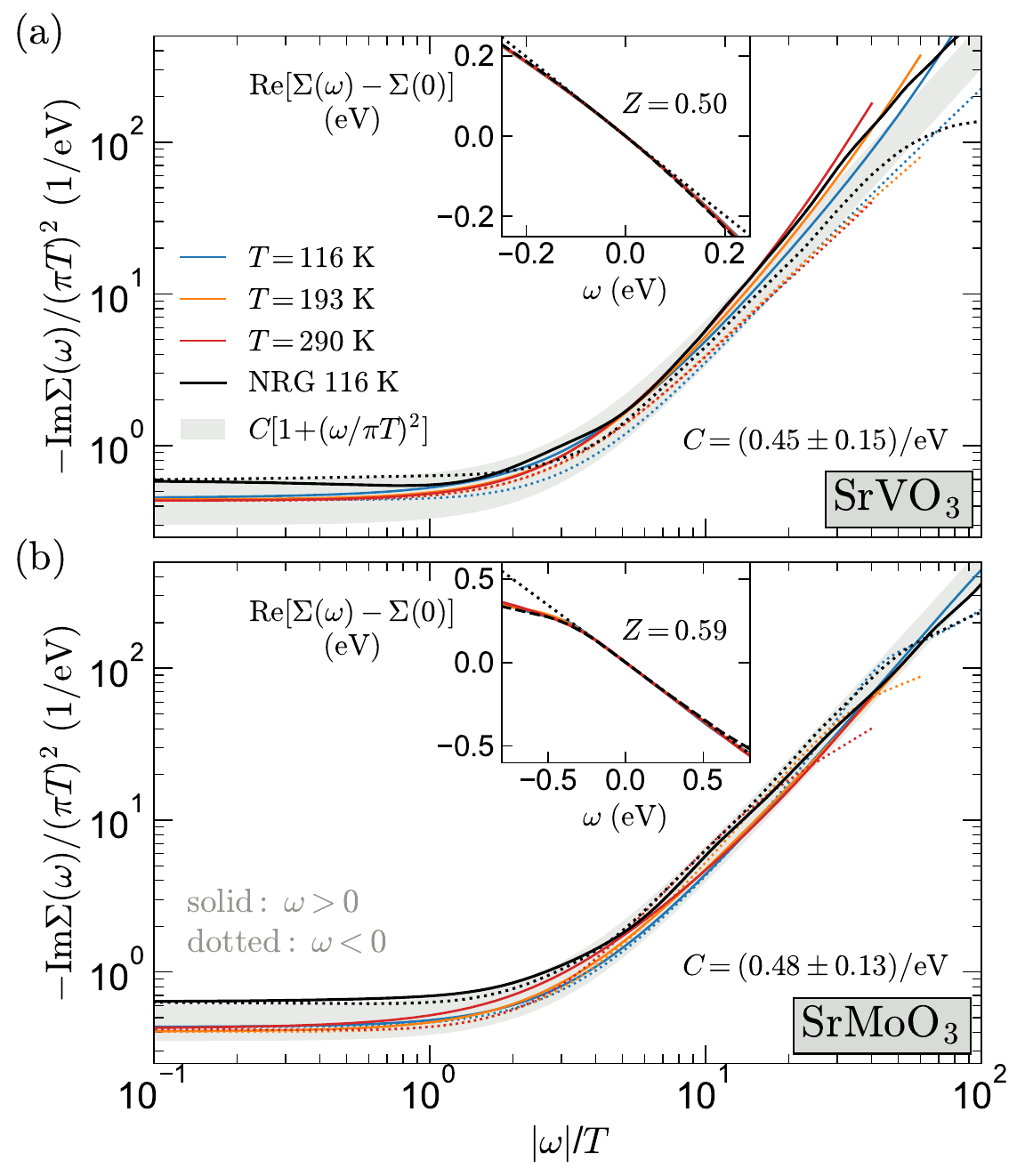}
\caption{$-\mathrm{Im}\Sigma(\omega)/(\pi T)^2$ as a function of $|\omega|/T$ for SrVO$_3$ and SrMoO$_{3}$ on a log-log scale. Solid (dotted) lines denote $\omega > 0$ ($\omega < 0$). Colored (black) lines are obtained with QMC (NRG). Shaded regions indicate a confidence region for the FL collapse to $C[\omega^{2}/(\pi T)^2 + 1]$. Inset shows $\mathrm{Re}\Sigma(\omega,T)-\mathrm{Re}\Sigma(0,T)$ compared with $1 - 1/Z=\partial_{\omega}\mathrm{Re}\Sigma(\omega,T)|_{\omega=0}$.}
\label{fig:ImSig_w}
\end{figure}

Importantly, the QMC+Pad\'e results for $\omega \gtrsim T$ are stable with respect to the AC parameters \cite{CompanionPaperLong} and in very good agreement with NRG. Hence, using the self-energies in this regime along with the known FL scaling~\eqref{eq:IMSig_FL} provides the most robust way to determine $C$. Figure~\ref{fig:ImSig_w} shows $-\text{Im}\Sigma(\omega,T)/(\pi T)^2$ vs.\ $\vert\omega\vert/T$ for $T=116$, 193, and $290$ K. As predicted by the FL relation~\eqref{eq:IMSig_FL}, the curves from different $T$ approximately collapse onto the universal form $C[\omega^2/(\pi  T)^2+1]$, establishing that these temperatures are below $T^{\text{FL}}$ for both materials \footnote{This is independently confirmed by our $T=0$ NRG results for the spin susceptibility (see \cite{CompanionPaperLong}).}. The resulting $C$ values are (the error bar/shaded region is chosen to span the spread of the curves in Fig.~\ref{fig:ImSig_w}) $0.45\pm 0.15\,\mathrm{eV}^{-1}$ for \svo{} \footnote{Note that this number is slightly larger than 0.33 eV$^{-1}$ reported in Ref.~\citenum{Abramovitch2024}; we show in \cite{CompanionPaperLong} that this discrepancy comes solely from the method of extracting $A$ from $\Sigma$} 
and $0.48\pm0.13\,\mathrm{eV}^{-1}$ for \smo{}.
Note that the presence of el-ph interactions in experiments requires measurements at very low $T$, while we can explore FL behavior in a broader $T$ range in our calculations involving only el-el interactions.

The insets in Fig.~\ref{fig:ImSig_w} show the real part of the self-energy, related to the quasiparticle weight $Z$ as
\begin{equation}
\label{eq:RESig_FL}
1 - 1/Z = \partial_\omega \text{Re}\Sigma(\omega,T)|_{\omega=0} .
\end{equation}
From both QMC and NRG, we find $Z \approx 0.50$ for \svo{} and $Z \approx 0.59$ for \smo{} (for all $T$ considered in Fig.~\ref{fig:ImSig_w}).

\begin{figure}
\includegraphics[width=\linewidth]{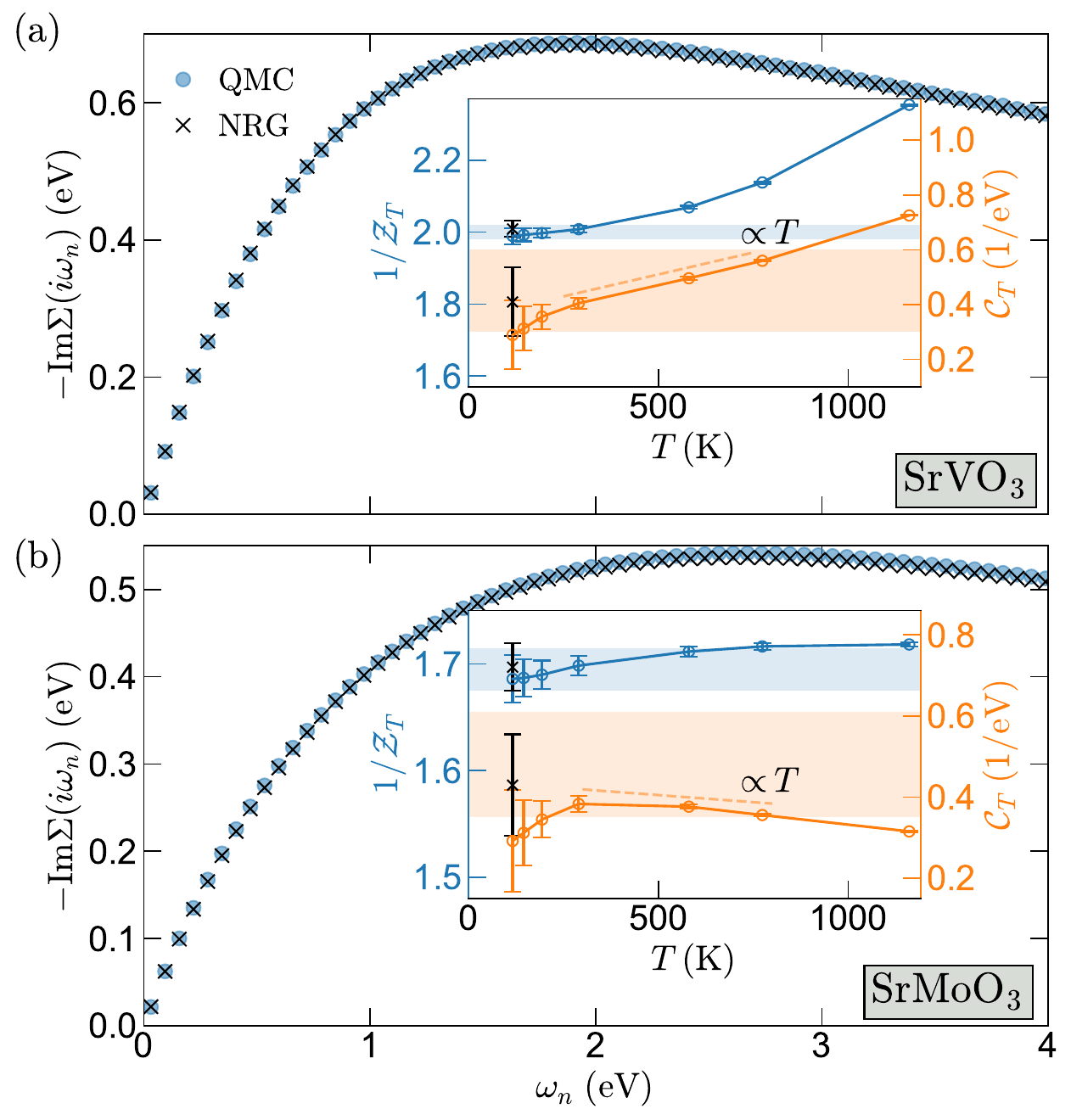}
\caption{$-\mathrm{Im}\Sigma(i\omega_{n})$ at $T = 116$~K for SrVO$_3$ and SrMoO$_3$, 
from QMC and NRG. Insets show $1/\mathcal{Z}_T$ and $\mathcal{C}_T$ computed from Eq.~\eqref{eq:SigMats_FL_formulas} for decreasing $T$. Error bars 
show how, by way of example, an error of $0.7$~meV on 
$\mathrm{Im}\Sigma(i\omega_{n})$
propagates to $1/\mathcal{Z}_T$ and $\mathcal{C}_T$.
Shaded regions indicate the range of values estimated from our real-frequency analysis in Fig.~\ref{fig:ImSig_w}.}
\label{fig:ImSig_iw}
\end{figure}

As an additional confirmation of these results, we perform a similar procedure on the Matsubara axis. To this end, we translate the FL forms~\eqref{eq:IMSig_FL} and \eqref{eq:RESig_FL} to Matsubara frequencies $i\omega_n \!=\! i(2n \!+\! 1)\pi T$, $n \!\in\! \mathbb{Z}$,
\begin{equation}
\label{eq:SigMats_FL}
\text{Im}\Sigma(i\omega_n,T) 
\simeq
(1 - 1/Z) \omega_n + \text{sgn}(\omega_n) C (\omega_n^2-\pi^2 T^2 )
.
\end{equation}
To avoid the problematic $i\omega\rightarrow i 0^+$ extrapolation, we can solve for the two parameters $Z$ and $C$ using the first two Matsubara points. We therefore define~\footnote{To simplify notations, the explicit dependence of 
$\Sigma$ on $T$ is omitted in this equation. Note that, when performing a linear fit to the first two Matsubara data points, the slope yields $1-1/Z + 4 C \pi T$, but the offset corresponds to $-4C \pi^2T^2$, hence a factor of $4$ times $\mathrm{Im}\Sigma(\omega=0,T)$. The factor 8 in the denominator of the estimator of $C$ used in the main text properly accounts for this.}
\begin{equation}
\label{eq:SigMats_FL_formulas}
1 - \frac{1}{\mathcal{Z}_T} = \frac{\text{Im}\Sigma(i\pi T)}{\pi T}
, \ \
\mathcal{C}_T = \frac{\text{Im}\Sigma(3i\pi T)-3\text{Im}\Sigma(i\pi T)}{8\pi^2 T^2}
.
\end{equation}
It is easy to see that, at low $T$, $\mathcal{Z}_T = Z + \mathcal{O}(T^2/T^2_{\text{FL}})$ \cite{Chubukov2012} and $\mathcal{C}_T = C + \mathcal{O}(T/T_{\text{FL}})$. The main panels of Fig.~\ref{fig:ImSig_iw} show $\text{Im}\Sigma(i\omega_n,T)$, demonstrating the agreement between QMC and NRG. The insets show how $\mathcal{Z}_T$ and $\mathcal{C}_T$ converge with decreasing $T$ to the values determined from our real-frequency analysis---although the expression for $\mathcal{C}_T$ becomes unstable to noise in $\mathrm{Im}\Sigma$ at low $T$.

Using Eq.~(\ref{eq:rho_FL}) to convert $C$ to $A$ coefficients (in units of $10^{-5}\mu\Omega\,$cm/K$^2$) gives $3.6 \pm 1.2$ for \svo{} and $2.2 \pm 0.6$ for \smo{} (see SM \cite{SM} Table~S1). We plot these results in the insets of Fig.~\ref{fig:exp}. For \svo{}, there is remarkable agreement with the low-$T$ $T^2$ regime of the highest-quality thin films, indicating the importance of el-el scattering in this regime \cite{Abramovitch2024,Brahlek2024}. For the \smo{} single-crystal data, the true low-$T$ behavior is not fully resolved, but appears to be consistent with a transition to a $T^2$ regime given by our $A$ coefficient. The fact that the $A$ coefficients of both materials are similar indicates that the significant difference in RT resistivity between the two oxides is not due to different degrees of electron correlation.

FL materials are often categorized by their so-called Kadowaki--Woods ratio~\cite{Rice1968,Kadowaki1986,Miyake1989,Hussey2005,Jacko2009} (KWR) $A/\gamma^2$. We compute the specific-heat coefficient $\gamma$ via the density of states at the Fermi level $D(\epsilon_{\text{F}})$ and $Z$,
\begin{equation}
\gamma=\frac{\pi^2 D(\epsilon_{\text{F}})}{3 Z}
, \ \ 
D(\epsilon_{\text{F}})=2 \sum_{m}\int_{\text{BZ}} \frac{d^dk}{(2\pi)^d} \delta(\epsilon_{\text{F}}-\epsilon_{m\textbf{k}})
.
\end{equation}
The results for the molar quantity $\gamma_0$ \cite{Hussey2005} in units of mJ/(\text{mol}$\,$K$^2$) are 7.9 for \svo{} and 7.5 for \smo{} (see SM \cite{SM}), in good agreement with the reported experimental values: 8.2 for \svo{} \cite{Inoue1998} and 7.9 for \smo{} \cite{Nagai2005}. Hence, the KWR in units of $\mu\Omega\,$cm$\,$mol$^2$K$^2$/J$^2$ is 0.6 for \svo{} and 0.4 for \smo{}. The KWR we find for \svo{} is more than an order of magnitude below the value of 10 reported for single crystals \cite{Inoue1998}. Our results thus cause the KWR of \svo{} to fall off the trend of other correlated oxides \cite{Hussey2005} (e.g., Sr$_3$Ru$_2$O$_7$ \cite{Behnia2022_2}), but puts it, along with \smo{}, in good agreement with elemental transition metals (e.g., Re or Fe) \cite{Rice1968}. This is suggestive of the fact that \svo{} and \smo{} are rather moderately correlated and consistent with the argument of Ref.~\citenum{Miyake1989} that weakly correlated systems (where $T^{\text{FL}}$ is comparable to $\epsilon_{\text{F}}$) would have a KWR similar to the transition metals.

In strongly correlated materials like UPt$_3$ \cite{Joynt2002} and cuprates \cite{Rullier2001,Rullier2003,Juskus2024}, the $A$ coefficient is basically independent of impurity concentration. However, for the moderately correlated \svo{} and \smo{}, there appears to be a correlation between the reported values of $A$ and $\rho_0$ (or anticorrelation between $A$ and RRR). Whether this is an experimental issue or indicates a breakdown of the simple picture in which impurities and disorder contribute a $T$-independent scattering rate via Matthiessen's rule is an important topic for further investigation. To settle this question, it is crucial to determine the origin of the discrepancy between single crystals and thin films in these materials, motivating further investigation on high-quality single crystals in the low-$T$ regime.

\begin{figure}
    \centering
    \includegraphics[width=0.85\columnwidth]{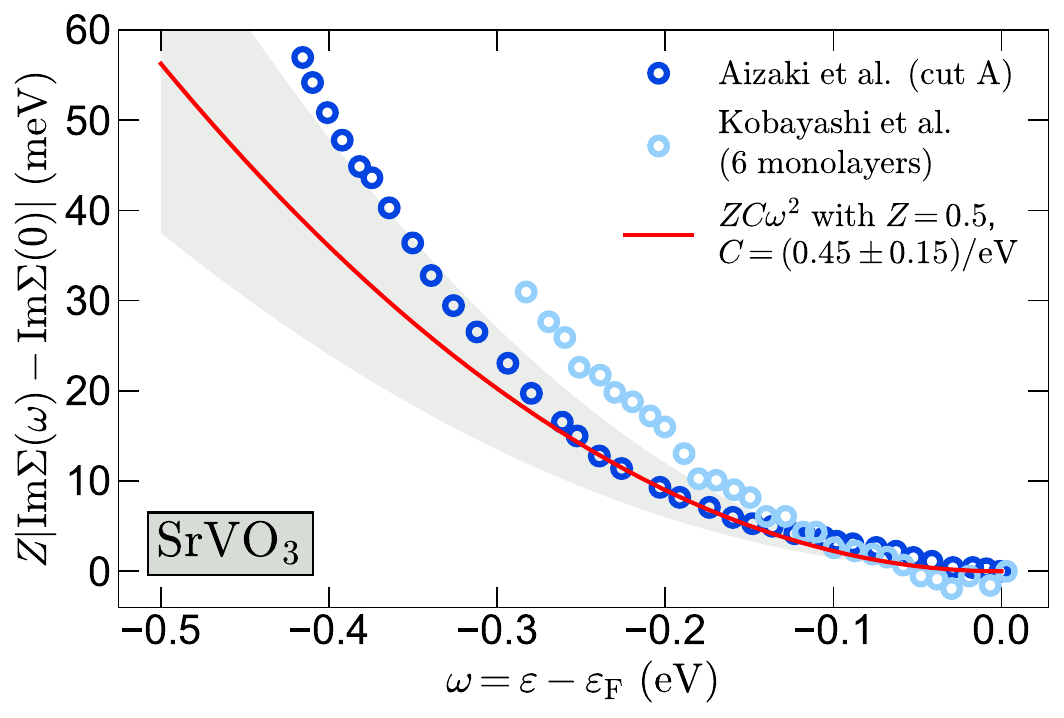}
    \caption{Comparison between 
    DMFT self-energy and ARPES momentum distribution curves (MDCs) for \svo{}, the latter taken from Aizaki \textit{et al.} \cite{aizaki2012} and Kobayashi \textit{et al.} \cite{kobayashi2017} (both measured below $20$~K). 
    We use
    $2Z|\text{Im}\Sigma(\omega)| \approx \hbar v_{\text{F}}^* \Delta k$, where $\Delta k$ is the MDC FWHM, and we extracted 
    the renormalized Fermi velocity as 
    $\hbar v_{\text{F}}^* = 0.54$~eV$\cdot$\AA{} from ARPES~\cite{aizaki2012}.}
    \label{fig:arpes-self-energy}
\end{figure}

Finally, we comment on fundamental implications of our work for transport theory. Our calculations use the \textit{single-site} DMFT approximation, where vertex corrections to the conductivity are absent due to the locality of the self-energy and vertex~\cite{Khurana1990}. However, vertex corrections were shown to be important in specific regimes, such as low-dimensional systems \cite{Brown2019,Vucicevic2019,Vranic2020,Vucicevic2023} and the low-density limit \cite{Mu2022,Mu2024}. In the low-density limit, phase-space considerations limit the possibility of Umklapp scattering, and it was shown in Ref.~\citenum{Mu2024} that $|\mathrm{Im}\Sigma|$ is severely underestimated by DMFT in this regime (note, though, that the case of a cylindrical Fermi surface, relevant to the present oxides, still needs to be explored along similar lines). In contrast to systems with very small Fermi surfaces like doped SrTiO$_3$ \cite{Lin2015_2,vandermarel_2011}, Umklapp scattering \textit{is} possible for \svo{}. In Fig.~\ref{fig:arpes-self-energy}, we compare our DMFT self-energy with angle-resolved photoemission spectroscopy (ARPES). The excellent agreement at low frequencies suggests that the possible underestimation of $|\mathrm{Im}\Sigma|$ does not apply here and provides a confirmation, independent of transport measurements, that single-site DMFT yields reliable results for the FL behavior of \svo{}. Nevertheless, the effect of vertex corrections to $\rho$ remains to be explored.

In summary, we showed that single-site DFT+DMFT can accurately describe the FL properties of \svo{} and \smo{}, two materials with moderate electron correlations. For \svo{}, we obtain an $A$ coefficient in agreement with the low-$T$ value for the best thin-film samples. Clarifying the experimental situation regarding the discrepancy with single-crystal measurements in this context is desirable. For \smo{}, our $A$ coefficient is lower than the fit to the experimental data. However, the experimental low-$T$ behavior is not fully resolved, and future experiments may reveal a lower-$T$ el-el-dominated $T^2$ regime as in \svo{}. This work thus serves as an impetus for more extensive transport measurements and computations for moderately correlated materials to elucidate fundamental aspects of transport theory. Specifically, an important extension of this work is to appropriately include \textit{non-local} correlation effects (GW \cite{Aryasetiawan1998}, cluster extensions of DMFT \cite{Maier2005,Rohringer2018}, etc.) and understand the relevance of vertex corrections. Experimentally, the crossover between the el-ph and el-el dominated regimes should be explored in spectroscopy alongside with transport. Overall, our work demonstrates the importance of synergetic developments in \textit{ab-initio} theory and materials synthesis/characterization to advance the understanding of transport in quantum materials.

\textit{Acknowledgments}---%
We are grateful to Philip Allen, Kamran Behnia, Ga\"{e}l Grissonnanche, Nigel Hussey, Andy Millis, Jernej Mravlje, Cyril Proust, and Louis Taillefer for useful discussions. The authors thank Naoki Shirakawa for sending us the data of Ref.~\citenum{Nagai2005}. C.E.D.\ and J.L.H.\ acknowledge support from the National Science Foundation under Grant No.~DMR-2237674. F.B.K.\ acknowledges funding from the Ministerium f\"ur Kultur und Wissenschaft des Landes Nordrhein-Westfalen (NRW-R\"uckkehrprogramm). The Flatiron Institute is a division of the Simons Foundation. The data that support the findings of this article are openly available \cite{Letter_data}.

\bibliography{bibfile}

\clearpage

\thispagestyle{empty}

\setcounter{equation}{0}
\setcounter{page}{1}
\setcounter{section}{0}
\setcounter{secnumdepth}{2} 
\renewcommand{\theequation}{S\arabic{equation}}
\renewcommand{\thepage}{S\arabic{page}}
\renewcommand{\thesection}{S\arabic{section}}

\title{Supplemental Material: \\ Fermi-liquid $T^2$ resistivity: Dynamical mean-field theory meets experiment}

\date{\today}
\maketitle

\onecolumngrid

\section{\label{sec:details}Computational details}
We compute the Kohn--Sham bands for cubic SrVO$_{3}$ ($a=3.863$ \AA{}) and SrMoO$_{3}$ ($a=4.007$ \AA{}) using the \textit{Vienna ab-initio simulation package} (VASP)~\cite{Kresse:1993bz,Kresse:1996kl,Kresse:1999dk}. Using Wannier90~\cite{wannier90}, we downfold the transition-metal $t_{2g}$-like bands onto a Wannier basis, which defines the impurity problem. We solve the quantum impurity problem in imaginary time using the QMC CT-HYB solver built on top of the TRIQS software stack~\cite{aichhorn_dfttools_2016, parcollet_triqs_2015, Seth2016274} with \texttt{solid\_dmft} as the driver of the DMFT loop~\cite{Merkel2022}, or in real frequencies using NRG (we exploit U(1) charge, SU(2) spin, and SO(3) orbital symmetries utilizing the MuNRG package \cite{Lee2021,Lee2017,Lee2016} and the QSpace tensor library \cite{Weichselbaum2012a,Weichselbaum2012b,Weichselbaum2020}). Both the QMC and NRG solvers use the same Wannier Hamiltonian as input. A Hubbard--Kanamori Hamiltonian including spin-flip and pair-hopping terms governs the interactions with $U = 4.5$~eV and $J = 0.65$~eV for \svo{}~\cite{Lechermann2006} and $U = 3.07$~eV and $J = 0.31$~eV for \smo{}~\cite{LeeHand2021,Hampel2021}. For additional computational details, see \cite{CompanionPaperLong}. See \cite{Sekiyama2004,Pavarini2004,Nekrasov2006,Maiti2006} for early works on \svo{} or, e.g., \cite{Kaufmann2021} for a list of references.

\section{\label{app:units}Units and values of DFT properties}
Restoring the prefactor $\hbar e^2$ for the transport function $\Phi$, the latter has units $\hbar e^2 \mathrm{cm}^{-3} (\mathrm{eV}\,\mathrm{cm}/\hbar)^2 \mathrm{eV}^{-1}$. Recognizing $e^2/\hbar$ as the conductance quantum, this immediately simplifies to $\mathrm{eV}/(\Omega \, \mathrm{cm})$. The precise values are
\begin{equation}
\text{\svo{}:} \,
\Phi(\epsilon_{\text{F}}) \!=\!  
\frac{2.247\,\mathrm{eV}}{\mathrm{m}\Omega\,\mathrm{cm}} 
, \ \
\text{\smo{}:} \,
\Phi(\epsilon_{\text{F}}) \!=\! 
\frac{3.930\,\mathrm{eV}}{\mathrm{m}\Omega\,\mathrm{cm}} 
.
\end{equation}
It follows that $A$ has units $(\Omega \, \mathrm{cm})/\mathrm{eV}^2$. Restoring a prefactor $k_B^2$, one obtains $A$ in units $(\Omega \, \mathrm{cm})/K^2$ and $\rho \sim A T^2$ in units $\Omega \, \mathrm{cm}$. The density of states $D$ as defined in the main text has units $\mathrm{eV}^{-1}\,\mathrm{cm}^{-3}$. However, for the molar specific-heat coefficient, the result is multiplied by the volume of the unit cell $V_{\text{uc}}$. The relevant values thus are
\begin{equation}
\text{\svo{}:} \,
V_{\text{uc}} D(\epsilon_{\text{F}}) = \frac{1.685}{\mathrm{eV}} 
, \ \
\text{\smo{}:} \,
V_{\text{uc}} D(\epsilon_{\text{F}}) = \frac{1.886}{\mathrm{eV}} 
.
\end{equation} 

\clearpage

\begin{table*}[h]
\renewcommand*{\arraystretch}{1.4} 
\caption{Fermi-liquid $T^2$ coefficient $A$ (fitted up to 300 K and at low-$T$, i.e., up to 20 K for \svo{} and 50 K for \smo{}), extrapolated zero-temperature resistivity $\rho_0$, and residual-resistivity ratio RRR from experimental measurements compared to the DMFT result for $A$ of this work. Dashes indicate that the fitting was not possible from the digitized data. Values in square brackets are those explicitly reported in the reference.}
\label{tab:svo_smo_exp}
\begin{ruledtabular}
\centering
\begin{tabular}{cc|c|cccc}
&&& \multicolumn{2}{c}{$A$ ($10^{-5}\mu\Omega\cdot$cm/K$^2$)} && 
\\
&&& Fit to 300 K & low-$T$ fit &$\rho_0$ ($\mu\Omega\cdot$cm) & RRR  
\\ \hline
\multirow{13}{*}{\bf{\svo{}}} &\multirow{4}{*}{Single crystal} & Berry \textit{et al.} \cite{berry2022} &526.8 [520.0] &-- &176.6 [172.4] &3.5 [3.6]
\\
&& Reyes Ardila \textit{et al.} \cite{Reyes2000} & 81.6 [83.0] & -- &[11.3] & 6.8 [6.0]
\\
&& Inoue \textit{et al.} \cite{Inoue1998} & 78.8  &[42.1]\footnotemark& [6.3] & 11.0
\\
&& Chamberland \textit{et al.} \cite{Chamberland1971} & 31.46  & -- & 0.47 & 58.94
\\ \cline{2-7}
&\multirow{9}{*}{Thin film} & Fouchet \textit{et al.} \cite{Fouchet2016} & 22.3 [22.1] & 34.2 & 27.0 [27.0] & 1.7 [2.0] 
\\
&& Xu \textit{et al.} \cite{xu2019} & 31.2 [27.7]& -- &11.3 
&3.4  
\\
&& Mirjolet \textit{et al.} \cite{mirjolet2021} & 37.3 [30.0]&7.0  &3.7 & 9.7 [9.6]
\\
&& Shoham \textit{et al.} \cite{Shoham2020} & 32.8 [33.5]& -- & 2.8 [2.9] &  11.1 [11.2]
\\
&& Roth \textit{et al.} \cite{Roth2021} &30.5& 11.8 & 1.4 [1.3] & 20.4 [21.0] 
\\
&& Zhang \textit{et al.} \cite{Zhang2016} & 35.6&7.9 & 0.3 & 95.9 
\\
&& Brahlek \textit{et al.} (2015) \cite{Brahlek2015} & 32.7 & 4.9 & 0.2 [0.2]  & 122.6 [125] 
\\
&& Ahn \textit{et al.} \cite{Ahn2022} & 30.3 & 6.7& 0.2 &126.2 [130]
\\
&& Brahlek \textit{et al.} (2024) \cite{Brahlek2024} &22.4 & 4.2  & 0.1 [0.1] & 185.7 [195.0]
\\ \cline{2-7}
&\multicolumn{2}{c|}{\textbf{DMFT}} & \multicolumn{2}{c}{$
\bm{3.6 \pm 1.2} 
$} & & 
\\
\hline\hline
\multirow{6}{*}{\bf{\smo{}}} & \multirow{5}{*}{Thin film} & Wang \textit{et al.} \cite{Wang2001} & $48.0$ &50.6& 83.7 & 1.3 
\\
&& Cappelli \textit{et al.}~\cite{Cappelli2022} & $37.0$ [$37.7\pm7$] &27.3& 39.3 [39.3]   & 1.6 [1.5] 
\\
&& Lekshmi \textit{et al.}~\cite{Lekshmi2005}& $74.3$ &93.3& 49.1 [35.7]  & 2.0 [1.9] 
\\
&& Radetinac \textit{et al.} \cite{Radetinac2016} & $27.4$ &25.7&11.6  &2.3
\\ \cline{2-7}
& Single crystal &Nagai \textit{et al.} \cite{Nagai2005} & 
$7.1$ [7.0]\footnotemark &
7.0 & 
0.35 [0.35] & 
14 [14]
\\ \cline{2-7}
& \multicolumn{2}{c|}{\textbf{DMFT}} & \multicolumn{2}{c}{$
\bm{2.2 \pm 0.6} 
$} &&
\\
\end{tabular}
\footnotetext[1]{Here, the authors used a fit up to 300 K, but including an el-ph $T^5$ contribution in the fit. Thus, $A$ reflects the low-$T$ behavior.}
\footnotetext[2]{Here, the high-$T$ fit is done up to 150 K instead of 300 K. Our low-$T$ fit is done from 10 K to 50 K and yields the same $\rho_0$ and $A$.}
\end{ruledtabular}
\end{table*}

\end{document}